\documentclass[aps,pra,onecolumn,groupedaddress,showpacs]{revtex4}

\usepackage{amssymb,amsmath}
\usepackage{graphicx}

\newcommand{\om}{\omega}

\newcommand{\pa}{\partial}

\begin{document}

\title{Two-component breather solution of the nonlinear wave equation}

\author{G. T. Adamashvili}
\affiliation{Technical University of Georgia, Kostava str.77, Tbilisi, 0179, Georgia.\\ email: $adamash77@gmail.com$ }

\begin{abstract}
A nonlinear wave equation which describes different nonlinear effects in various fields of research, was considered. In two particular cases, this equation was reduced to the Sine-Gordon equation and the Born-Infeld equation. Using the slowly varying envelope approximation and the generalized perturbative reduction method, the nonlinear wave equation was transformed to coupled nonlinear Schr\"odinger equations for auxiliary functions. An explicit analytical solution of a nonlinear wave equation in the form of a two-breather molecule was obtained. One breather oscillated with the sum, and the other with the difference of frequencies and wave numbers. The obtained solution coincides with the solutions of the two-breather molecule found in a number of well-known equations from different areas of physics. It is shown that in a particular case of the small amplitude waves, a solution in the form of a two-breather molecule for the nonlinear Klein-Gordon equation coincides with the vector $0\pi$ pulse of the self-induced transparency which is presented under less stringent conditions compared to the same solution of this equation obtained earlier.
\vskip+0.2cm
\emph{Keywords:} Generalized perturbation reduction method, Two-breather molecule, Born-Infeld equation, Sine-Gordon equation, Nonlinear Klein-Gordon equation.
\end{abstract}

\pacs{05.45.Yv, 02.30.Jr, 52.35.Mw}

\maketitle

\section{Introduction}

At the propagation of nonlinear waves, one of the most important and interesting phenomena is the formation of nonlinear solitary waves of a stationary form. Although these waves have been studied and observed for a very long time, interest in them does not fade away. This is because they express the most fundamental property of strongly nonequilibrium states of nonlinear systems when the parameters characterizing such states remain unchanged. Although nonlinear solitary waves occur in various fields of physics, for waves of a completely different nature and physical functions characterizing them, for example, the strength of the electric field of optical waves, the tensor of deformation for sound waves, the displacement of the surface from the undisturbed state for the water waves and others, nevertheless their properties are identical. The nonlinear solitary waves which arise in different nonlinear physical phenomena in various systems, are described by means of the various nonlinear partial differential equations. These equations include the Korteweg de Vries equation, the Boussinesq equation, the Hirota equation, the nonlinear Schr\"odinger equation, the Benjamin-Bona-Mahony equation, the Bloch-Maxwell system of equations and many others [1-9].

Among the nonlinear partial differential equations which have solutions in the form of nonlinear solitary waves, one can single out the well-known Sine-Gordon equation and the Born-Infeld equation. Each of these equations describes nonlinear solitary waves in different areas of research. For example, the Sine-Gordon equation describes the geometry of surfaces with Gaussian curvature, the Josephson transition, dislocations in crystals, waves in ferromagnets associated with the rotation of magnetization, the properties of elementary particles and the nonlinear phenomena in the various fields of physics: optics, acoustics, plasma physics, semiconductor quantum dots, graphene, optical and acoustical metamaterials, and others [1-4]. The Born-Infeld equation is used to describe the interaction of electromagnetic waves in nonlinear electrodynamics, various phenomena in the field theory, the theory of strings, and some of the atomic experiments [1, 10-14]. These equations have been studied in sufficient detail by various mathematical methods and their solutions are well known.

Although each of these two equations describes a fairly wide range of absolutely different phenomena in various fields of physics, nevertheless, both of these equations are nonlinear Maxwell wave equations but with different nonlinear terms. Consequently arises the natural question: is it possible to describe all the above effects connected with these two equations in a unified way, by means of one more general equation?

In order to answer this question we consider the following more general nonlinear wave equation
\begin{equation}\label{h}
\frac{\partial^{2} U}{\partial t^{2}}-C \frac{\partial^{2} U}{\partial z^{2}} =-\alpha_{0}^{2} \sin U -  A (\frac{\partial U}{\partial t})^{2}   \frac{\partial^{2} U}{\partial z^{2}} - \sigma
(\frac{\partial U}{\partial z})^{2}   \frac{\partial^{2} U}{\partial t^{2}}     + B \frac{\partial U}{\partial t}\frac{\partial U}{\partial z} \frac{\partial^{2} U}{\partial t   \partial z},
\end{equation}
or in the dimensionless form
\begin{equation}\label{h1}\nonumber
 \frac{\partial^{2} U}{\partial t^{2}}-\frac{\partial^{2} U}{\partial z^{2}} =- \sin U- (\frac{\partial U}{\partial t})^{2}   \frac{\partial^{2} U}{\partial z^{2}}-
(\frac{\partial U}{\partial z})^{2}   \frac{\partial^{2} U}{\partial t^{2}}     + 2 \frac{\partial U}{\partial t}\frac{\partial U}{\partial z} \frac{\partial^{2} U}{\partial t   \partial z},
\end{equation}
where $U(z,t)$ is a real function of space coordinate $z$ and time $t$ and represents the wave profile, while  $\alpha_{0}^{2},\;$ $A,\;B,\;$ $ C,$ and $\sigma $ are the real constants.

Eq.(1) in particular cases goes into the Sine-Gordon equation and the Born-Infeld equation.
Indeed, when the condition  $A=B=\sigma=0 $  is fulfilled, Eq.(1) is transformed into the Sine-Gordon equation
\begin{equation}\label{sg}
\frac{\partial^{2} U}{\partial t^{2}}-C \frac{\partial^{2} U}{\partial z^{2}} =-\alpha_{0}^{2} \sin U.
\end{equation}

This nonlinear equation was analyzed by means of inverse scattering transform [2, 4], by which it is possible to obtain the complete solution of Eq.(2) in the form of the nonlinear solitary waves.

Among nonlinear solitary waves, the nonlinear waves have a relatively low amplitude for which
\begin{equation}\label{t}
U<<1
\end{equation}
are considered quite often. Under the condition Eq.(3), the Sine-Gordon equation (2) is reduced to the nonlinear Klein-Gordon equation [1]
\begin{equation}\label{kg}
 \frac{\partial^{2} U}{\partial t^{2}}-C \frac{\partial^{2} U}{\partial z^2} =-\alpha_{0}^{2} U +\frac{\alpha_{0}^{2}}{6} U^{3} -\mathcal{O}( U^5).
\end{equation}

In case when the condition $\alpha_{0}^{2}=0$ is fulfilled, Eq.(1) is reduced to the Born-Infeld equation
\begin{equation}\label{bi}
\frac{\partial^{2} U}{\partial t^{2}}-C \frac{\partial^{2} U}{\partial z^{2}} =-  A (\frac{\partial U}{\partial t})^{2}   \frac{\partial^{2} U}{\partial z^{2}} - \sigma
(\frac{\partial U}{\partial z})^{2}   \frac{\partial^{2} U}{\partial t^{2}}     + B \frac{\partial U}{\partial t}\frac{\partial U}{\partial z} \frac{\partial^{2} U}{\partial t   \partial z}.
\end{equation}

This is a nonlinear modification of the Maxwell wave equation, which includes the so-called Born-Infeld nonlinearity.

Sometimes $ 2+1$ (two space coordinates and time) dimensional version of the Born-Infeld equation also has been investigated [15].

Under the condition Eq.(3), the nonlinear wave equation (1) is reduced to the form
\begin{equation}\label{hk}
\frac{\partial^{2} U}{\partial t^{2}}-C \frac{\partial^{2} U}{\partial z^{2}} = -  A (\frac{\partial U}{\partial t})^{2}   \frac{\partial^{2} U}{\partial z^{2}} - \sigma
(\frac{\partial U}{\partial z})^{2}   \frac{\partial^{2} U}{\partial t^{2}}     + B \frac{\partial U}{\partial t}\frac{\partial U}{\partial z} \frac{\partial^{2} U}{\partial t   \partial z}-\alpha_{0}^{2} U +\frac{\alpha_{0}^{2}}{6} U^{3} -\mathcal{O}( U^5).
\end{equation}

When considering nonlinear solitary waves, there are two different types of them: single-component (scalar) and two-component (vector) solitary nonlinear waves. These waves can be formed both in different and in the same physical system. Their properties and formation conditions are different. A single-component wave propagates in a medium without changing its profile. Completely different is a two-component nonlinear wave, which is a bound state of two wave packets having the same speeds and directions of propagation, and in general cases, different oscillation frequencies. These wave packets can have both parallel polarizations or mutually perpendicular polarizations, for example, for waveguide modes. Such a two-component wave is a vector breather or a two-breather molecule. In the process of propagation, the breathers that make up the molecule interact with each other and exchange energy.

Some effects cannot be described using single-component nonlinear waves and it becomes necessary to use the concept of a breather molecule [16-18]. The most striking example is the effect of self-induced transparency, one of the main pulses of which is precisely a special type of the two-breather molecule, which consists of two connected breathers with the same polarizations, one of the breathers oscillates with the sum, and the second with the difference in frequencies and wave numbers. Such a two-component solitary nonlinear wave was first considered at the nonlinear coherent interaction of the optical wave with the system of the resonance atoms and is called a vector $0\pi$ pulse of self-induced transparency [19-22].

Various mathematical approaches are used to analyze the properties of solitary nonlinear waves. Among them, the perturbative reduction method for analysis of the single-component nonlinear waves very often meet [23, 24]. This method uses one complex auxiliary function and two real parameters, which is enough to study single-component nonlinear waves but not enough to study two-component nonlinear waves. To study two-component nonlinear waves, a generalized perturbative reduction method was developed (see, for instance [9, 19-22] and references therein) which uses two complex functions and eight real parameters which are enough for the analysis of two-component nonlinear waves.

In the theory of self-induced transparency, the role of the second derivatives of the strength of the electric field of the wave with respect to the spatial coordinate and time in the Maxwell wave equation was studied using various mathematical methods. In particular, using the perturbative reduction method it was found that the second derivatives lead to only small quantitative corrections to the values of the parameters of the pulses of self-induced transparency (see, for instance [25, 26]).

Using the generalized perturbative reduction method, in the theory of self-induced transparency qualitatively new result was obtained. Namely, thanks to this method it was shown that in the Maxwell wave equation the second derivatives of the strength of the electric field of the pulse with respect to the spatial coordinate and time the determining role play for an adequate study of the effect of the self-induced transparency. To take into account the terms of the second derivative in the Maxwell wave equation was allowed to prove that one of the main pulses of self-induced transparency is the vector $0\pi$ pulse (two-breather molecule), but not the single-component scalar $0\pi$ pulse, as this was previously believed. The reason for this is that in the study of self-induced transparency in the Maxwell wave equation, the second derivatives of the strength of the electric field of the pulse with respect to the spatial coordinate and time were neglected, or their influence in the process of the formation of the pulses of self-induced transparency was investigated by using the standard perturbative reduction method, or by means of some other mathematical methods which intended for analysis of the single-component waves [19-22, 25-34].

Although the vector $0\pi$ pulse of self-induced transparency, which is a special type of two-breather molecule, was first noticed for optical nonlinear waves, later, using the generalized perturbative reduction method, the same two-breather molecules were discovered for waves of a different nature described by other nonlinear partial differential equations, such as are the Boussinesq-type equation, the Benjamin-Bona-Mahony equation, and the Hirota equation [7, 9, 35-37].

Taking into account that a two-breather molecule in which one breather oscillates with the sum, and the second with the difference in frequencies and wave numbers, qualitatively changes the physical picture of self-induced transparency that cannot be described within the framework of single-component nonlinear waves, as well as the fact that such a two-breather molecule arises in completely different areas of physics for a function of a different nature, allows concluding that existence of this type of a two-breather molecule, i.e. type of the vector $0\pi$ pulse of self-induced transparency, expresses the rather general property of matter. Thus, its further study for new equation (6) is relevant and reasonable.

The main purpose of this work is to use the generalized perturbative reduction method to obtain a solution to the nonlinear wave equation (6) in the form of the two-breather molecule, in which one breather oscillates with the sum and the second with the difference in frequencies and wave numbers.

The rest of this paper is organized as follows: In Section II, we consider the nonlinear wave equation (6) for slowly varying complex envelope functions. Using the generalized perturbative reduction method, we transformed the obtained equation into coupled nonlinear Schr\"odinger equations for auxiliary functions. In Section III, we present the explicit analytical solution of a nonlinear wave equation (6) in the form of a two-breather molecule Eq.\eqref{vb}. Finally, in Section IV, we discuss the obtained results.

\vskip+0.5cm

\section{The generalized perturbative reduction method}

We considered a solitary nonlinear wave with the carrier frequency $\omega$ and wave number $k$ propagating along the positive $z$-axis. To study solitary nonlinear waves, the value of the width of pulse $T$ is of great importance. We considered pulses whose width satisfies the condition $T>>1/\omega$, i.e. the pulse duration is much longer than the inverse frequency of the carrier wave. For such a wave, we can use the slowly varying envelope approximation, and for real function $U$, use the expansion [5, 38]
\begin{equation}\label{eq2}
U(z,t)=\sum_{l=\pm1}\hat{u}_{l}(z,t) Z_l,
\end{equation}
where $Z_{l}= e^{il(k z -\omega t)}$  is the exponential fast oscillating function, and $\hat{u}_{l}$ is the slowly varying complex envelope function, which satisfies inequalities
\begin{equation}\label{swa}\nonumber
 \left|\frac{\partial \hat{u}_{l}}{\partial t}\right|\ll\omega |\hat{u}_{l}|,\;\;\;\;\;\;\;\;\;\;\;
 \left|\frac{\partial \hat{u}_{l}}{\partial z }\right|\ll k|\hat{u}_{l}|.
\end{equation}
For the reality of $U$, it is supposed that the expression $ \hat{u}_{+1}= \hat{u}^{*}_{-1}$  is valid.

The  nonlinear wave equation (6) can be presented as
\begin{equation}\label{nlin}
\frac{\pa^{2} U}{\pa t^2} -C \frac{\pa^{2} U}{\pa z^{2}}=-\alpha_{0}^{2} U +N(U),
\end{equation}
where
\begin{equation}\label{nhk}
N(U)= -  A (\frac{\partial U}{\partial t})^{2}   \frac{\partial^{2} U}{\partial z^{2}} - \sigma
(\frac{\partial U}{\partial z})^{2}   \frac{\partial^{2} U}{\partial t^{2}}     + B \frac{\partial U}{\partial t}\frac{\partial U}{\partial z} \frac{\partial^{2} U}{\partial t   \partial z}+\frac{\alpha_{0}^{2}}{6} U^{3} -\mathcal{O}( U^5)
\end{equation}
is the nonlinear term of the equation.

Substituting  Eq.(7) into (8), we obtained the following dispersion relation
\begin{equation}\label{dis}
{\omega}^{2}= C k^{2}+\alpha_{0}^{2},
\end{equation}
and the rest of the equation (8) in the form
\begin{equation}\label{shk}
\sum_{l=\pm1}Z_l (-2il\omega \frac{\partial \hat{u}_{l}}{\partial
t}- 2ilkC \frac{\partial \hat{u}_{l}}{\partial z}+\frac{\partial^{2} \hat{u}_{l}}{\partial t^{2}} -C
\frac{\partial^{2} \hat{u}_{l}}{\partial z^{2}})=N(\hat{u}_{l}).
\end{equation}
where $N(\hat{u}_{l})$ is the nonlinear part of Eq.(9), expressed by means of the complex envelope function $\hat{u}_{l}$.

In order to consider the two-component nonlinear wave solution of Eq.(6), we used the generalized perturbative reduction method developed in Refs.[9, 19-22], which made it possible to transform the nonlinear wave equation (11) for the function $\hat{u}_{l}$ to the coupled nonlinear Schr\"odinger equations for auxiliary functions.

Following this method, the complex envelope function  $\hat{u}_{l}$ can be represented as
\begin{equation}\label{gprm}
\hat{u}_{l}(z,t)=\sum_{\alpha=1}^{\infty}\sum_{n=-\infty}^{+\infty}\varepsilon^\alpha
Y_{l,n} f_{l,n}^ {(\alpha)}(\zeta_{l,n},\tau),
\end{equation}
where $\varepsilon$ is a small parameter,
$$
Y_{l,n}=e^{in(Q_{l,n}z-\Omega_{l,n}
t)},\;\;\;\zeta_{l,n}=\varepsilon Q_{l,n}(z-v_{{g;}_{l,n}} t),
$$$$
\tau=\varepsilon^2 t,\;\;\;
v_{{g;}_{l,n}}=\frac{\partial \Omega_{l,n}}{\partial Q_{l,n}}.
$$

It is assumed that the complex auxiliary functions $f_{l,n}^{(\alpha)}$ and oscillating real parameters $\Omega_{l,n}$ and $Q_{l,n}$ satisfy the inequalities for any $l$ and $n$
\begin{equation}\label{rtyp}\nonumber\\
\omega\gg \Omega_{l,n},\;\;k\gg Q_{l,n},\;\;\;
\end{equation}
$$
\left|\frac{\partial
f_{l,n}^{(\alpha )}}{
\partial t}\right|\ll \Omega_{l,n} \left|f_{l,n}^{(\alpha)}\right|,\;\;\left|\frac{\partial
f_{l,n}^{(\alpha )}}{\partial \eta }\right|\ll Q_{l,n} \left|f_{l,n}^{(\alpha )}\right|.
$$

Substituting Eq.(12) into Eq.(11), we obtained the nonlinear wave equation for auxiliary function $f_{l,n}^{(\alpha)}$ in the following form:
\begin{equation}\label{eqz}
\sum_{l=\pm1}\sum_{\alpha=1}^{\infty}\sum_{n=\pm 1}\varepsilon^\alpha Z_{l} Y_{l,n}[W_{l,n}
+\varepsilon J_{l,n} - \varepsilon^2 i l h_{l,n}  \frac{\partial }{\partial \tau}
-\varepsilon^{2} Q^{2} H_{l,n}\frac{\partial^{2} }{\partial \zeta^{2}}+O(\varepsilon^{3})]f_{l,n}^{(\alpha)}=N(\hat{u}_{l}),
\end{equation}
where
\begin{equation}\label{cof}
W_{l,n}=- 2 n l\omega \Omega_{l,n}  + 2  n  l k Q_{l,n} C - \Omega^{2}_{l,n}+C  Q_{l,n}^{2},
$$
$$
J_{l,n}=2i Q_{l,n} [l \omega  v_{{g;}_{l,n}}   -l k C   + n \Omega_{l,n}  v_{{g;}_{l,n}}   -C n Q_{l,n}],
$$
$$
h_{l,n}=2(\omega + ln \Omega_{l,n}),
$$
$$
H_{l,n}=   C- v_{{g;}_{l,n}}^{2}.
\end{equation}

Following the standard procedure and equating to zero, the terms with the same powers of $\varepsilon$, from Eq.(13), we obtained a series of equations. In the first order of $\varepsilon$, we found a connection between the parameters  $\Omega_{l,n}$ and $Q_{l,n}$. When
\begin{equation}\label{fo1}
2 ( C k Q_{\pm1, \pm1} -\omega \Omega_{\pm1, \pm1}) - \Omega^{2}_{\pm1, \pm1} + C  Q_{\pm1, \pm1}^{2}=0,
\end{equation}
then $ f_{\pm1, \pm1}^{(1)}\neq0$  and when
\begin{equation}\label{fo2}
2 ( C k Q_{\pm1, \mp1} -\omega \Omega_{\pm1, \mp1}) + \Omega^{2}_{\pm1, \mp1} - C  Q_{\pm1, \mp1}^{2}=0,
\end{equation}
then $ f_{\pm1, \mp1}^{(1)}\neq0$.

From Eq.(14), in  the second order of $\varepsilon$, we obtained the equation
$J_{\pm1, \pm1}=J_{\pm1, \mp1}=0$ and the expression
\begin{equation}\label{v}
v_{{g;}_{l,n}}=C \frac{  k + l n Q_{l,n} }{ \omega   +l n  \Omega_{l,n} }.
\end{equation}

Next, we considered the nonlinear term $N(\hat{u}_{l})$ of the nonlinear wave equation (13). Substituting Eqs.(12) and (7) into Eq.(9) for the nonlinear term of the nonlinear wave equation, we obtained
\begin{equation}\label{nonlz}
  \varepsilon^{3}\;Z_{+1}[(\tilde{q}_{+} | f_{+1,+1}^ {(1)}|^{2} + \tilde{r}_{+} | f_{+1,-1}^ {(1)}|^{2} ) f_{+1,+1}^ {(1)}Y_{+1,+1}
   + (\tilde{q}_{-} |f_{+1,-1}^ {(1)}|^{2} + \tilde{r}_{-} |f_{+1,+1} ^ {(1)}|^{2} ) Y_{+1,-1} f_{+1,-1}^ {(1)}]
 \end{equation}
and plus terms proportional to $ Z_{-1}$. In Eq.(18) we use the notations
\begin{equation}\label{qr}
\tilde{q}_{\pm}=\frac{\alpha_{0}^{2}}{2}+\mathfrak{q}_{\pm},
$$
$$
\tilde{r}_{\pm}=\alpha_{0}^{2}+\mathfrak{r}_{\pm},
\end{equation}
where
$$
\mathfrak{q}_{\pm}=(A +\sigma -B )(\om \pm \Omega_{\pm } )^{2}  (k\pm  Q_{\pm} )^{2},
$$
$$
\mathfrak{r}_{\pm}=2[ A  (k\pm  Q_{\pm} )^{2}  (\om \mp\Omega_{\mp} )^{2}    +  \sigma  ( \omega \pm \Omega_{\pm }  )^{2}     (k \mp Q_{\mp}  )^{2}
-B     (  \om + \Omega_{+}  )   (  \om -\Omega_{-} ) (  k+ Q_{+}  )  (   k- Q_{-} )].
$$

From Eqs.(13) and (18), in the third order of $\varepsilon$, we obtained the system of nonlinear equations
\begin{equation}\label{2eq}
  i \frac{\partial f_{+1,+1}^{(1)}}{\partial \tau} + Q_{+}^{2} \frac{H_{+1,+1} }{h_{+1,+1}} \frac{\partial^2 f_{+1,+1}^{(1)}}{\partial \zeta_{+1,+1} ^2}+(\frac{\tilde{q}_{+}}{ h_{+1,+1}}   | f_{+1,+1}^ {(1)}|^{2} + \frac{\tilde{r}_{+}}{ h_{+1,+1}} | f_{+1,-1}^ {(1)}|^{2} ) f_{+1,+1}^ {(1)}=0,
$$$$
   i \frac{\partial f_{+1,-1 }^{(1)}}{\partial \tau} + Q_{-}^{2} \frac{H_{+1,-1} }{h_{+1,-1}} \frac{\partial^2 f_{+1,-1 }^{(1)}}{\partial \zeta_{+1,-1}^2}+(\frac{\tilde{q}_{-}}{h_{+1,-1}}  |f_{+1,-1}^ {(1)}|^{2} +\frac{\tilde{r}_{-}}{h_{+1,-1}} |f_{+1,+1} ^ {(1)}|^{2} )  f_{+1,-1}^ {(1)}=0.
 \end{equation}

\vskip+0.5cm
\section{Two-breather molecule of the nonlinear wave equation }

After transforming back to the space coordinate $z$ and time $t$, from Eqs.(20), we obtained the coupled nonlinear Schr\"odinger equations for the auxiliary functions $\Lambda_{\pm}=\varepsilon  f_{+1,\pm1}^{(1)}$ in the following form:
\begin{equation}\label{pp2}
i (\frac{\partial \Lambda_{\pm}}{\partial t}+v_{\pm} \frac{\partial  \Lambda_{\pm}} {\partial z}) + p_{\pm} \frac{\partial^{2} \Lambda_{\pm} }{\partial z^{2}}
+q_{\pm}|\Lambda_{\pm}|^{2}\Lambda_{\pm} +r_{\pm} |\Lambda_{\mp}|^{2} \Lambda_{\pm}=0,
\end{equation}
where
\begin{equation}\label{OmQ}
v_{\pm }= v_{g;_{+1,\pm 1}}=C \frac{  k \pm  Q_{\pm} }{ \omega   \pm   \Omega_{\pm } },,\;\;\;\;\;\;\;\;\;\;\;\;\;\;p_{\pm}=\frac{ C- v_{\pm}^{2} }{2(\omega \pm \Omega_{\pm})},
$$
$$
q_{\pm}=\frac{ \tilde{q}_{\pm}}{2(\omega \pm \Omega_{\pm})},\;\;\;\;\;\;\;\;\;\;\;\;\;\;\;\;\;
r_{\pm}=\frac{ \tilde{r}_{\pm}}{2(\omega \pm \Omega_{\pm})},
$$
$$
 \Omega_{+}=\Omega_{+1,+1}= \Omega_{-1,-1},\;\;\;\;\;\;\;\;\;\;\;\;\;\;\;\;\;\;\;\;\;\;\;\;\;  \Omega_{-}= \Omega_{+1,-1}= \Omega_{-1,+1},
$$
$$
 Q_{+}=Q_{+1,+1}= Q_{-1,-1},\;\;\;\;\;\;\;\;\;\;\;\;\;\;\;\;\;\;\;\;\;\;\;\;\;  Q_{-}= Q_{+1,-1}= Q_{-1,+1}.
\end{equation}

The solution of Eq.(21) is given by [19, 20, 22]
\begin{equation}\label{ue1}
\Lambda_{\pm }=\frac{A_{\pm }}{b T}Sech(\frac{t-\frac{z}{V_{0}}}{T}) e^{i(k_{\pm } z - \omega_{\pm } t )},
\end{equation}
where $A_{\pm },\; k_{\pm }$ and $\omega_{\pm }$ are the real constants, and $V_{0}$ is the velocity of the nonlinear wave. We assume that
$k_{\pm }<<Q_{\pm }$  and $\omega_{\pm }<<\Omega_{\pm }.$

Combining Eqs.(7), (12) and (23), we obtained the two-breather molecule solution to the nonlinear wave  equation (6) in the following form:
\begin{equation}\label{vb}
U(z,t)=\mathfrak{A} Sech(\frac{t-\frac{z}{V_{0}}}{T})\{  \cos[(k+Q_{+}+k_{+})z -(\omega +\Omega_{+}+\omega_{+}) t]
$$$$
+(\frac{p_{-}q_{+}-p_{+}r_{-}} {p_{+}q_{-}- p_{-}r_{+}})^{\frac{1}{2}} \cos[(k-Q_{-}+k_{-})z -(\omega -\Omega_{-}+\omega_{-})t]\},
\end{equation}
where $\mathfrak{A}$ is the amplitude of the nonlinear pulse. The expressions for  the parameters $k_{\pm }$ and $\omega_{\pm }$  are given by
\begin{equation}\label{omk}
k_{\pm }=\frac{V_{0}-v_{\pm}}{2p_{\pm}},
$$$$
\omega_{+}=\frac{p_{+}}{p_{-}}\omega_{-}+\frac{V^{2}_{0}(p_{-}^{2}-p_{+}^{2})+v_{-}^{2}p_{+}^{2}-v_{+}^{2}p_{-}^{2}
}{4p_{+}p_{-}^{2}}.
\end{equation}

\vskip+0.5cm

\section{Conclusion}

We investigated the two-breather molecule solution of the nonlinear wave equation (6) when the slowly varying envelope approximation Eq. \eqref{eq2} was valid.
The nonlinear pulse with the width $T>>\Omega_{\pm }^{-1}>>\omega^{-1}$ was considered.

Using the generalized perturbation reduction method Eq.(12), Eq.(6) is transformed to the coupled nonlinear Schr\"odinger equations (21) for the function $\Lambda_{\pm 1}$. As a result, the two-component nonlinear pulse oscillating with the sum $\omega +\Omega_{+}$ $(k+Q_{+})$ and difference $\omega -\Omega_{-}$ ($k-Q_{-}$) of the frequencies (wave numbers) Eq.(24) (two-breather molecule) was obtained. The dispersion relation and the relations between oscillating parameters $\Omega_{\pm}$ and $Q_{\pm}$ were determined from Eqs.(10), (15) and (16). The parameters of the nonlinear pulse from Eqs.(17), (19), (22) and (25) were determined.

Eq.(6) can describe all effects, which can be considered by means of the nonlinear Klein-Gordon equation Eq.(4) and the Born-Infeld equation Eq.(5) separately. Eqs.(24) are also solutions of these equations, and all obtained results are valid for these two equations.

In the special case when $\alpha_{0}^{2}=0$, Eq.(6)  is reduced to the Born-Infeld equation Eq.(5). The dispersion law Eq.(10) for the Born-Infeld equation is reduced to the form ${\omega}^{2}= C k^{2},$ and the parameters of the wave have the form $\tilde{q}_{\pm}=\mathfrak{q}_{\pm}$ and $\tilde{r}_{\pm}=\mathfrak{r}_{\pm}.$

In the second special case when the condition $A=B=\sigma=0 $  is fulfilled, Eq.(6) is reduced to the nonlinear Klein-Gordon equation Eq.(4). The dispersion law Eq.(10) is valid, and the parameters of the wave have the form $\tilde{q}_{\pm}=\frac{\alpha_{0}^{2}}{2}$ and $\tilde{r}_{\pm}=\alpha_{0}^{2}$.

Although the Sine-Gordon equation (2) and the nonlinear Klein-Gordon equation (4) arise in a number of physical fields and applied mathematics, there is one more very important effect of self-induced transparency, the simplified version of which is based on the Sine-Gordon equation, where the function $U$ is the optical pulse envelope [38]. Indeed, the nonlinear coherent interaction of an optical pulse with resonant optical atoms is governed by the Bloch-Maxwell equations [27, 38]. When the Rabi frequency of the wave is real and the longitudinal and transverse relaxations are ignored, these equations are reduced to the Sine-Gordon equation [1-4], and for small amplitude waves when $U<<1$, the Sine-Gordon equation is reduced to the nonlinear Klein-Gordon equation (4).

Ref.[19] shows that, under the conditions of self-induced transparency, Eq.(4) has a two-component vector $0\pi$ pulse solution. One component oscillates with the sum $w+\Omega_{+}$ ($\kappa+Q_{+}$), and the other with the difference  $w-\Omega_{-}$ ($\kappa-Q_{-})$ between the frequencies (wave numbers) in the region of the parameters $w$ and $\kappa$ which that are two or three orders lower than the carrier wave frequency $\omega$ and wave number $k$. In other words, the expressions  $\omega/w$ and $k/\kappa$ are of orders $10^{2}\div10^{3}$. In this case, the conditions of the nonlinear wave existence
\begin{equation}\label{cn1}
\omega>>w>>\Omega_{\pm}>>T^{-1},\;\;\;\;\;\;\;\;\;\;\;\;\;\;\;\;\;\;k>>\kappa>>Q_{\pm}>>(V_{0}T)^{-1}
\end{equation}
are fulfilled. Here $w,$ $\kappa$,  $\Omega_{\pm}$ and $Q_{\pm}$ are the different oscillating parameters. The phase modulation is neglected.

Taking into account the phase modulation, the character of the wave process changes significantly. As shown in Refs.[20-22], the two-component vector $0\pi$ pulse solution of the Bloch-Maxwell equations oscillates with the sum and difference in frequencies in the region of the carrier wave frequency and wave number of the optical pulse. In this case, the conditions of formation of the two-component vector $0\pi$ pulse have a form
\begin{equation}\label{cn}
\omega>>\Omega_{\pm}>>T^{-1},\;\;\;\;\;\;\;\;\;\;\;\;\;\;\;\;\;\;\;\;\;k>>Q_{\pm}>>(V_{0}T)^{-1},
\end{equation}
that is considerably weaker than Eq.(26), as shown in Ref.[19].

From Eq.(6), in the special case when the condition  $A=B=\sigma=0 $ is fulfilled, Eq.(6) is transformed to the nonlinear Klein-Gordon equation (4), which has a solution in the form of Eq.(24) under the condition Eq.(27). When function $U$ is the area of the optical pulse envelope, the two-breather molecule Eq.(24) coincides with the vector $0\pi$ pulse of self-induced transparency [19-22].

We found that Eq.(24) is a solution of Eqs.(6) and (4) under the condition Eq.(27), as was obtained earlier in the case of the Bloch-Maxwell equations but without the phase modulation. In other words, in the present work, we obtained the solution in the form of the two-breather molecule or the vector $0\pi$ pulse of self-induced transparency Eq.(24)  for the Eq.(4)  under the less stringent conditions Eq.(27), compared to the solution of this equation obtained earlier in Ref.[19] under the condition Eq.(26).

The considered solution in the form of a two-breather molecule - one of the breathers oscillates with the sum and the second with the difference in frequencies Eq.(24) - shows that this nonlinear wave occurs in many completely different areas of research for various nonlinear partial differential equations. Therefore, it can be concluded that the existence of such a nonlinear solitary wave expresses a rather general property of matter, as well as this takes place for the soliton or the breather.

We considered the nonlinear wave equation (1) (Eq.(6)), which unified the Sine-Gordon equation (2) (the nonlinear Klein-Gordon equation (4)) and the Born-Infeld equation (5). A similar approach used earlier for the Hirota equation, which in one particular case reduced to a scalar nonlinear Schr\"odinger equation, and in another particular case, it transformed into a complex modified Korteweg de Vries equation. Notable, there was a difference; unlike the nonlinear wave equations (1) and (6), where the function $U(z,t)$ satisfying these equations was a real function, in the cases of the Hirota equation, the scalar nonlinear Schr\"odinger equation and the complex modified Korteweg de Vries equation, the solutions of these equations were found to be complex functions [7, 39, 40].

\vskip+2.5cm

\end{document}